\documentclass[aps, prl, twocolumn, superscriptaddress]{revtex4-2}
\usepackage{graphicx}
\usepackage{amsfonts}
\usepackage{float}
\usepackage{amsmath}
\usepackage{latexsym}
\usepackage{bm, bbold}
\usepackage{upgreek}
\newcommand{\tr}{\mathrm{tr}}

\begin{document}

\title{
Wilson loop and Wilczek-Zee phase from a non-Abelian gauge field
}

\date{\today}

\author{Seiji Sugawa}
    \affiliation{Joint Quantum Institute, 
    National Institute of Standards and Technology and University of Maryland,
    Gaithersburg, Maryland 20899-8424 USA}
    \affiliation{Institute for Molecular Science, National Institutes of Natural Sciences, Myodaiji, Okazaki 444-8585, Japan} 
    \affiliation{SOKENDAI (The Graduate University for Advanced Studies), Myodaiji, Okazaki 444-8585, Japan}
    \author{Francisco Salces-Carcoba}
    \altaffiliation[Present Address: ]{LIGO Laboratory, California Institute of Technology, MS 100–36, Pasadena, CA 91125, USA}
    \affiliation{Joint Quantum Institute, 
    National Institute of Standards and Technology and University of Maryland,
    Gaithersburg, Maryland 20899-8424 USA}

\author{Yuchen Yue}
    \affiliation{Joint Quantum Institute, 
    National Institute of Standards and Technology and University of Maryland,
    Gaithersburg, Maryland 20899-8424 USA}

\author{Andika Putra}
    \altaffiliation[Present Address: ]{The MathWorks, Inc., 3 Apple Hill Drive, Natick, Massachusetts 01760 USA}
    \affiliation{Joint Quantum Institute, 
    National Institute of Standards and Technology and University of Maryland,
    Gaithersburg, Maryland 20899-8424 USA}

\author{I.~B. Spielman}
    \affiliation{Joint Quantum Institute, 
    National Institute of Standards and Technology and University of Maryland,
    Gaithersburg, Maryland 20899-8424 USA}

\begin{abstract}
    Quantum states can acquire a geometric phase called the Berry phase after adiabatically traversing a closed loop, which depends on the path not the rate of motion.
    The Berry phase is  analogous to the Aharonov-Bohm phase derived from the electromagnetic vector potential, and can be expressed in terms of an Abelian gauge potential called the Berry connection.
    Wilczek and Zee extended this concept to include non-Abelian phases --- characterized by the gauge independent Wilson loop--- resulting from non-Abelian gauge potentials.
    Using an atomic Bose-Einstein condensate, we quantum-engineered a non-Abelian SU(2) gauge field, generated by a Yang monopole located at the origin of a 5-dimensional parameter space.
    By slowly encircling the monopole, we characterized the Wilczek-Zee phase in terms of the Wilson loop, that depended on the solid-angle subtended by the encircling path: a generalization of Stokes' theorem.
    This observation marks the observation of the Wilson loop resulting from a non-Abelian point source.
\end{abstract}
\maketitle
\section{INTRODUCTION}

    The seemingly abstract geometry of a quantum system's eigenstates now finds application in fields ranging from condensed-matter and quantum information science to high-energy physics.
    The Berry curvature--- a geometric gauge field present for a single (non-degenerate) quantum state moving in any parameter space~\cite{Berry1984} --- is a prime observable associated with this geometry.   
    The Berry phase is the direct analogue to the Aharonov-Bohm phase for motion along a closed loop with the enclosed Berry curvature playing the role of magnetic field.
    Berry's curvature and phase have been measured in a variety of physical systems throughout physics and chemistry~\cite{Zak1989berry, Leek2007, Kendrick2015, Yuan2018, GeometricPhase_book}, and even have an analog for planetary-scale atmospheric waves~\cite{Delplace2017}.
    Monopoles, or conical intersections, singular points in the energy landscapes of a range of physical systems~\cite{DiracMonopole, Zhang2005_graphene, Duca2015_AB, Yuan2018, CI2020} where the curvature diverges,  play a crucial role in geometric effects, since particles encircling the singular point can acquire non-zero Berry phase.

    The Wilczek-Zee (W.-Z.) phase~\cite{WilczekZee1984} extends these ideas to include non-Abelian `operator-valued' geometric phases possible for adiabatically evolving systems with a degenerate subspace (DS).
    Initial nuclear magnetic resonance experiments~\cite{Tycko1987, Zwanziger1990} inspired holonomic quantum computation utilizing the W.-Z. phase to affect noise-resistant geometric quantum gates~\cite{Duan2001,Abdumalikov2013,zu2014,arroyo2014,Toyoda2013}.
    A non-Abelian phase has been also studied and characterized in ‘non-degenerate’ multi-band optical lattice on non-cyclic paths at the strong-force limit~\cite{Li2016}.
    Despite the universality of Wilczek and Zee's concept and the tremendous theoretical and experimental interest in synthetic non-Abelian gauge fields
    ~\cite{NonAbelianBO, SO3phase,Ruseckas2005, Osterloh2005, Goldman2009_Wilsonloop, monopole2009, Bermudez2010_honeycomb, Hauke2012, Phuc2015, Toyoda2013, leroux2018non,Yang1021}, 
    there has been no realistic cold-atom scheme for robust control of non-Abelian geometric phase in an adiabatic matter, nor a measurement of non-trivial gauge-independent Wilson loop on a closed path that characterizes the non-trivial non-Abelian geometric phase in cold-atom systems.
    
    Here, we observed and characterized the W.-Z. phase in an atomic Bose-Einstein condensate (BEC) as it underwent near-adiabatic motion in a five-dimensional (5D) parameter space with a non-Abelian Yang monopole at its origin~\cite{Yang1978, YangMonopole2018}.  
    Each point in this synthetic dimensional parameter space defined the Hamiltonian for four atomic hyperfine states, and the W.-Z. phase, governed by a non-Abelian SU(2) gauge field, described the adiabatic response within a spin-1/2 DS.
    We obtain the analog to Stokes' theorem, connecting W.-Z. phase to the solid-angle subtended by a closed path by characterizing the phase with gauge-independent Wilson loop (WL)~\cite{Wilson1974}.
    
	This geometric process, shown in Fig.~1\textbf{c}, can be viewed as moving a test particle in 5D around the Yang monopole, the source of the SU(2) gauge field~\cite{Yang1978, YangMonopole2018}. 
    The Yang monopole is a non-Abelian generalization of Dirac monopole~\cite{Ray2014_monopole}, and characterized by a non-zero second Chern number. 
    Our manuscript is organized as follows: (1) we introduce the essential physics of the WL, (2) describe our experimental setup.  We then (3) show the non-Abelian operator character of the W.-Z. phase factor using quantum process tomography, and (4) characterize it in terms of the gauge-independent WL.  Lastly, (5) we comment on extensions of these techniques to larger gauge groups, such as the ${\rm SU}(3)$ gauge group of the strong nuclear force.
    
    Although the usual Berry connection, the gauge potential associated with the Berry curvature, is gauge dependent~\cite{Berry1984}, both the Berry phase and Berry curvatures are gauge independent. 
    Specifically, they are invariant under the local gauge transformation $U=\mathrm{e}^{\mathrm{i}\Phi(\bm q)}$ for any choice of position-dependent phase $\Phi(\mathbf{q})$.
	In contrast, non-Abelian extensions of Berry's phase and curvatures need not be gauge invariant.
    The WL, defined as trace of the W.-Z. phase factor (non-Abelian holonomy) is a gauge-independent geometric quantity that reduces to the Berry phase for a single non-degenerate state~\cite{footnote1}.
    It was originally considered for the problem of quark-confinement~\cite{Wilson1974, Gils1981} and is often used in formulating gauge theories.
    In topological quantum computation, the WL describes braiding evolution of non-Abelian anyons~\cite{Kitaev2003, Pachos2012_TPQ}.
    Moreover, in crystalline systems -- including both conventional materials and synthetic quantum matter -- the eigenspectrum of the WL can characterize the topology of multiple Bloch bands~\cite{Alexandradinata2014_Wison, Li2016, newRef1, newRef2}.

    In the framework of differential geometry~\cite{Nakahara_book}, adiabatic motion is described in terms of fiber bundles, where the fibers represent the gauge degree of freedom. 
    As the state adiabatically evolves, a parallel transport condition sets the choice of basis state leading to a vertical lift along the fiber (Fig.~1\textbf{d}).
    After tracing out a closed loop $\mathcal{C}$ in space, the state will have evolved according to the unitary transformation
        \begin{equation}
            \hat{U}_\mathcal{C} = \mathcal{P}\exp\left(
            \mathrm{i} \int_\mathcal{C} \boldsymbol{\mathcal{\hat{A}}}_\mathbf{q} \cdot d\mathbf{q}
            \right),
            \label{holonomy}
        \end{equation}
    the W.-Z. geometric phase factor, i.e., the non-Abelian holonomy.
    Here $\mathcal{P}$ indicates that the exponential should be evaluated in a path-ordered manner and $\boldsymbol{\mathcal{\hat{A}}}_\mathbf{q}$ is a non-Abelian gauge field (non-Abelian Berry connection).
    The cyclic property of the trace makes the WL $W=\tr(\hat{U})$ manifestly gauge independent.
    Previous experimental work on a multi-band system in an optical lattice characterized the matrix elements of the non-Abelian holonomy along a non-cyclic path in crystal-momentum space and reconstructed the gauge-dependent Wilson line~\cite{Li2016}. 
    Similar to this work, we characterize the non-Abelian holonomy. However, we measured its process matrices instead to include potential imperfection in the analysis to prove the near-unity fidelity of our robust control, in addition to the reconstruction of the WL. 
    We demonstrate that a gauge-independent WL on a closed path can be fully tuned, in contrast to the previous study in multi-band system~\cite{Li2016}, where gauge-independent WL on a closed path was trivial. 
        
    \begin{figure*}[t!]
        \includegraphics[width=5in]{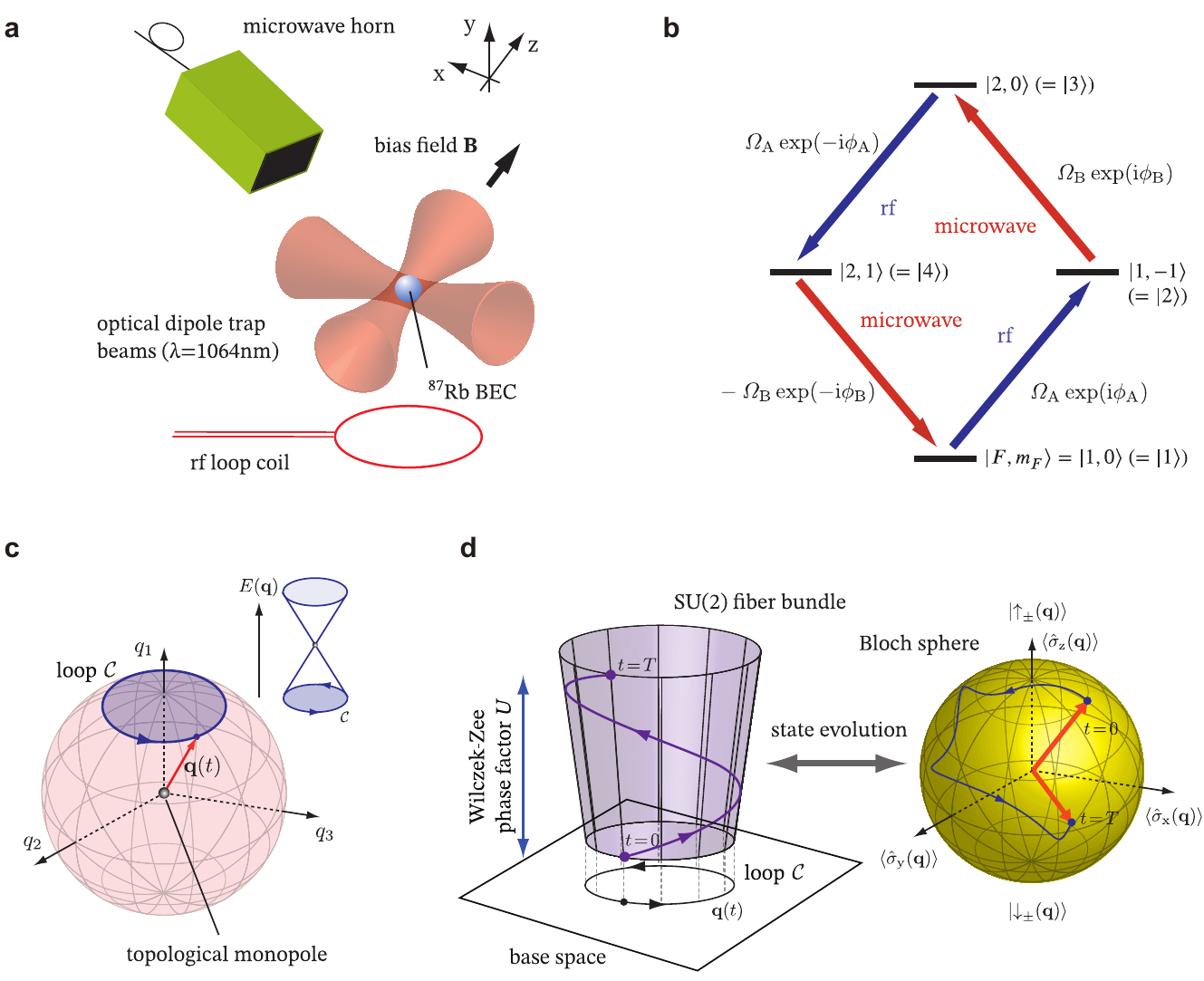}
        \caption{
        \textbf{{Experimental and conceptual schematic}}.
        \textbf{a} Experimental setup. A $^{87}$Rb BEC subject to uniform bias magnetic field was illuminated with rf and microwave fields which coupled its hyperfine ground states.
        \textbf{b} Cyclically coupled four-level system realized with hyperfine ground states of $^{87}$Rb.
         {The total phase of the four complex coupling is $\uppi$\cite{YangMonopole2018}}.
        \textbf{c} W.-Z. phase measurement.
        A test particle encircling the Yang monopole along a path $\mathcal{C}$ acquires a W.-Z. phase.
        (inset) Energy landscape. The synthesized Yang monopole is a singularity at four-fold degenerate point, where the energy gap $\Updelta\mathrm{E}=0$.	
        \textbf{d} Fiber-bundle description of the W.-Z. phase.  
        Left: the vertical lift along the SU(2) fiber bundle describes the W.-Z. phase.  
        Right:  The state evolution in our DS is represented by the trajectory of the Bloch vector.  
        The final state differs from the initial state by a factor of the holonomy.}
    \end{figure*}
    
    \begin{figure}
        \includegraphics[width=3.3in]{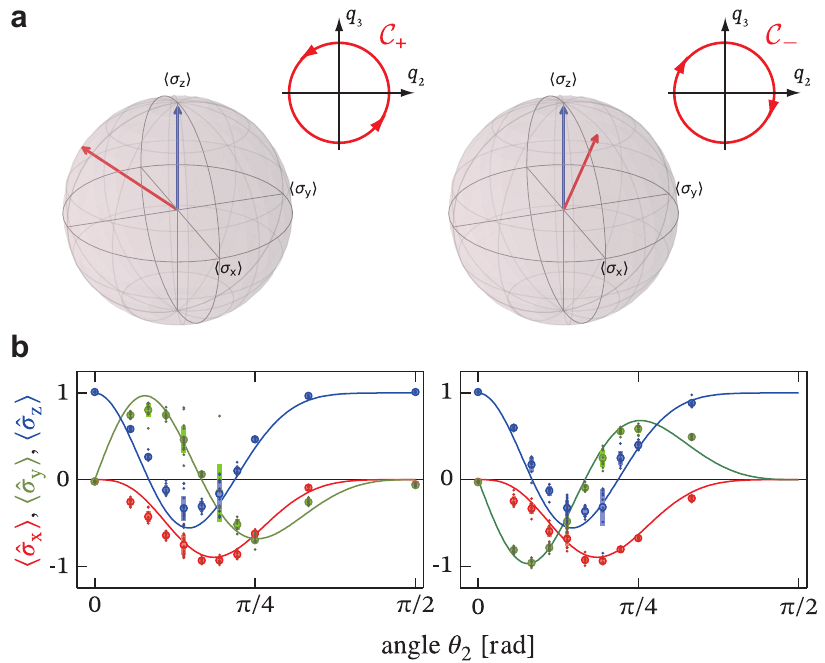}
        \caption{
        \textbf{{Acquisition of W.-Z. phase within the DS}}.
        \textbf{a} Bloch vector within the DS before (blue arrows) and after (red arrows) adiabatically following paths $\mathcal{C}_\pm$, which traced the same loop with opposite direction (inset).
        The initial Bloch vector was $\langle \hat{\boldsymbol{\upsigma}} \rangle=(0, 0, 1)$ and the path dependent final Bloch vectors were  $\langle \hat{\boldsymbol{\upsigma}}_+ \rangle=(-0.62(3), -0.70(5), 0.47(2))$ and $\langle \hat{\boldsymbol{\upsigma}}_- \rangle=(-0.68(1), 0.59(5), 0.40(5))$.
        The laboratory parameters are $\varOmega_{\mathrm A}=\varOmega_{\mathrm B}=h \times 1.4 \mathrm{kHz}$.
        \textbf{b} The angle dependence of the final Bloch vectors for paths with forward ramp (left) and reverse ramp (right). 
        {The dots are individual data and} the solid curves are the theory.
        {The error bars show the standard deviation of the data.}
        }
    \end{figure}
   
    \begin{figure}
        \includegraphics[width=3.3in]{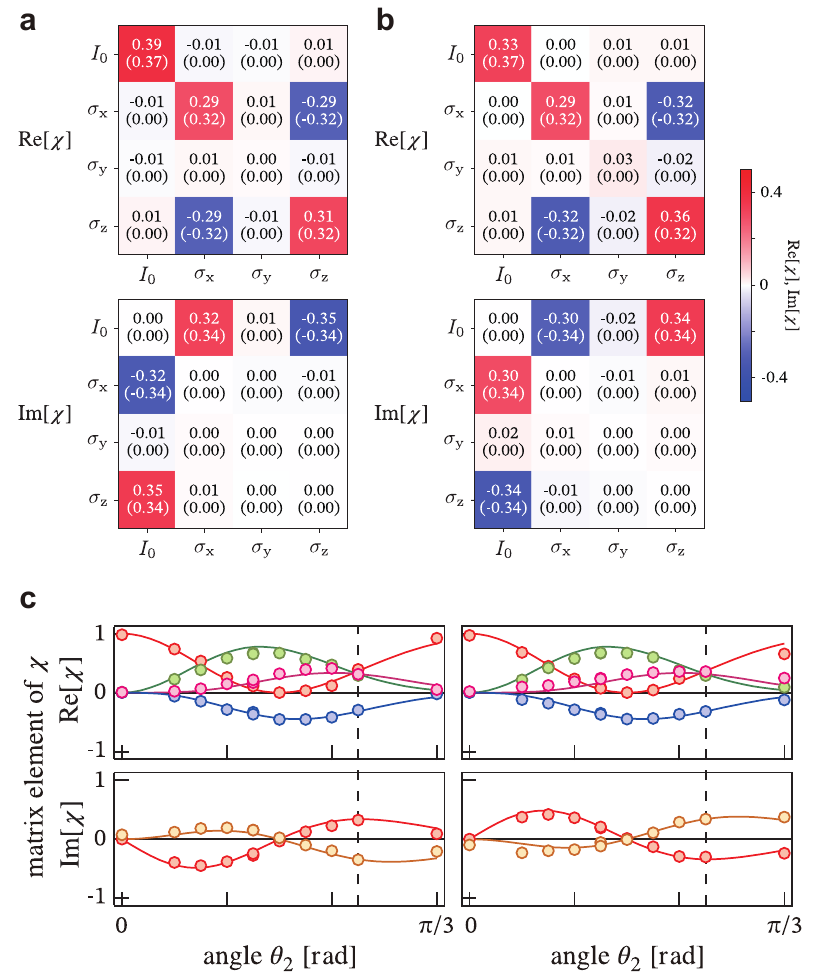}
        \caption{
        \textbf{{Quantum process tomography}}.
        \textbf{a,b} Reconstructed process matrices $\chi_{ij}$ of the non-Abelian holonomy $\hat{U}_{\mathcal C}$ for \textbf{a} forward and \textbf{b} reversed ramps at $\theta_2=\uppi/4$. The real part (Re[$\chi_{ij}$]) and the imaginary part (Im[$\chi_{ij}$]) are shown separately.
        Values in brackets are theoretical.
        The fidelity of the geometric process reached $\mathcal{F}_{\mathcal{C}_+}=0.98$ and $\mathcal{F}_{\mathcal{C}_-}=0.96$ for forward and reversed ramp, respectively.
        \textbf{c} Process matrix components ($\chi_{ij}$) for forward (left panels) and reverse (right panels) ramps for different W.-Z. phase realizations.
        The top panel shows the real parts 
        [$\chi_{\rm 00}$ (red), $\chi_{\rm xx}$ (green), $\chi_{\rm zz}$ (purple), $\chi_{\rm xz}$ (blue)] and 
        the bottom panels shows the imaginary parts 
        [$\chi_{\rm 0x}$ (red), $\chi_{\rm 0z}$ (orange)].
        We note that $\chi$ is Hermitian. 
        The solid curves are the theory.
        Matrix components that are constantly zero in the theory are not shown for clarity. 
        }
    \end{figure}
    
    \begin{figure}
        \includegraphics[width=3.3in]{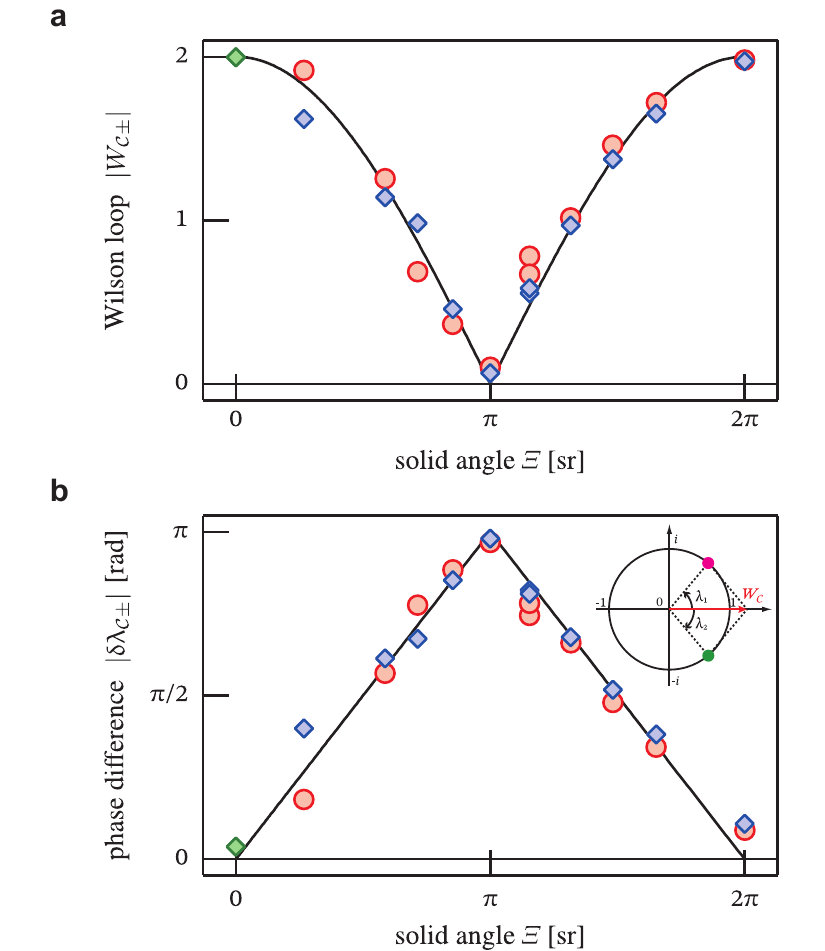}
        \caption{
        \textbf{{Wilson loop.}}
        \textbf{a} $|W_{\mathcal{C}_\pm}|$ for forward ($W_{\mathcal{C}_+}$, blue diamonds) and reverse ($W_{\mathcal{C}_-}$, red circles) ramps are plotted along with theory (solid curve).
        At $\varXi=0$ ($\theta_2=\uppi/2$), the circular paths reduce to a point, which we trivially measure with $T$=0~ms (green diamond).
        \textbf{b} The phase difference of the eigenvalues of W.-Z phase factor  $|\updelta\uplambda_{\mathcal{C}_\pm}|$ obtained from the WL measurement in \textbf{a}, with the same symbols.
        Inset: In the complex plane, the eigenvalues appear on the unit circle
        (pink and green points), and sum to $W_{\mathcal{C}}$ (red arrow).
        }
    \end{figure}
    
    \section{Results}
    \textit{Experimental setup}
    We prepared $^{87}{\rm Rb}$ BEC with $\approx 1 \times 10^5$ atoms in $|F,m_F\rangle=|1,-1\rangle$ in a crossed optical dipole trap formed by two horizontal 1064 nm optical trapping beams.
    We engineered a non-Abelian SU(2) gauge field with the BECs, using four $|F,m_F\rangle$ hyperfine ground states~\cite{YangMonopole2018}: $\{|1,0\rangle, |1,-1\rangle, |2,0\rangle, |2,1\rangle\}$ respectively labeled $\{|1\rangle, |2\rangle, |3\rangle, |4\rangle\}$.
    The 19.8 G bias magnetic field (with 2.5 ppm long-term stability), resolved the rf and microwave transition frequencies within the hyperfine states (Fig.~1\textbf{a}).  
    As shown in Fig.~1\textbf{b}, we coupled these states with rf and microwave fields parameterized by two Rabi frequencies $\varOmega_{{\rm A}}$ and $\varOmega_{{\rm B}}$ with phases $\phi_{{{\rm A}}}$ and $\phi_{{\rm B}}$.  We parameterize the coupling ratio $\varOmega_{\rm B}/\varOmega_{\rm A}=\tan\theta_2$ in terms of an angle $\theta_2$. The system then evolved according to the Hamiltonian
    \begin{equation}
        \hat{H}
        =-\frac{\hbar}{2}\sum_{i=1}^5 q_i \hat{\varGamma}_i,
        \label{Hamiltonian}
    \end{equation}
    expressed in terms of the reduced {Planck} constant $\hbar$, and the five Dirac gamma matrices $\hat{\varGamma}_i$.
    In addition, the vector $\mathbf{q}=(q_1, q_2, q_3, q_4, q_5)$ defines coordinates in a 5D parameter space, and is determined by laboratory parameters 
    $q_1=-\varOmega_{\rm B} \cos\phi_{\rm B}$,
    $q_2=-\varOmega_{\rm A} \cos\phi_{\rm A}$, 
    $q_3=-\varOmega_{\rm A} \sin\phi_{\rm A}$, 
    $q_4=\delta_{\rm Z}$, and
    $q_5=-\varOmega_{\rm B} \sin\phi_{\rm B}$.
    The detuning $\delta_{\rm Z}$, from the linear Zeeman shift, is set to zero throughout our measurement.
    The Hamiltonian can be represented by a 4-by-4 matrix by taking the four hyperfine states as the basis, where the diagonal part shows the detuning, and the non-zero off-diagonal elements shows the coupling between the hyperfine states.
    The resulting spectrum, insensitive to environmental noise such as magnetic field fluctuations, always consisted of a pair of two-fold degenerate energy manifolds with eigenstates $\{|{\uparrow}_{-} (\mathbf{q})\rangle, |{\downarrow}_{-}(\mathbf{q})\rangle\}$ for the ground state manifold  {(See Eq.~5)} and $\{|{\uparrow}_{+}(\mathbf{q})\rangle, |{\downarrow}_{+}(\mathbf{q})\rangle\}$ for the excited state manifold.
    Throughout this manuscript, the gap ($\Updelta\mathrm{E}( \mathbf{q})=\hbar |\mathbf{q}|=\hbar \sqrt{\varOmega_{\rm A}^2+\varOmega_{\rm B}^2} $) 
    is $h \times 2.0$ kHz, and was measured by inducing coherent Rabi-like oscillations between the eigenstates [see Methods].
    Due to the two-fold DS, the underlying gauge field $\boldsymbol{\mathcal{\hat{A}}}$ is non-Abelian and has SU(2) symmetry.
    When $\mathbf{q}$ is adiabatically changed along a closed path,
    the quantum state evolves within the subspace and acquires W.-Z. phases.

\textit{Wilczek-Zee phase}
    The consequence of the acquired W.-Z. phase can be experimentally captured by explicitly following an initial state as it evolves within the DS as a result of adiabatically moving $\mathbf{q}$ in parameter space.
    We demonstrated this by preparing an eigenstate 
    ($|{\uparrow}_{-}(\mathbf{q}_0)\rangle = 1/\sqrt{2}|1\rangle-1/2|2\rangle+1/2|4\rangle$) at $\mathbf{q}_0=
    (-\varOmega_{\rm B},-\varOmega_{\rm A},0,0,0)$ with $\theta_2=\mathrm{\uppi}/4$
    in the ground state manifold.
    After the state preparation, we linearly ramped the rf phase ($\phi_{\rm A}(t)=2\mathrm{\uppi} t/T$, where $T=2$ ms), tracing out a closed loop $\mathcal{C}_+$. 
    We performed state tomography within the DS to compare the initial and final states.
    
    The state within the DS is described by the Bloch vector on a Bloch sphere.
    The states before (blue) and after (red) the control sequence are shown in the left panel of Fig.~2\textbf{a}.
    The Bloch vector is rotated even though the final control parameters are the same as the initial ones, manifesting the operator content of the W.-Z. phase factor.
    This is striking in contrast with the Abelian Berry phase, which would leave the orientation on the Bloch sphere unchanged.

    Geometric phases depend on ramp-direction, however, for the Abelian case reversing the ramp direction along the same closed path simply inverts the sign of the phase. 
    For the non-Abelian case, this relation does not hold, indeed, the right panel of Fig.~2\textbf{a} shows the final state is quite different when the ramp is reversed.
    We denote motion along the same path in the reversed direction by $\mathcal{C}_-$.
    The trajectories can be varied by changing $\theta_2$, with initial eigenstate $|{\uparrow}_-(\mathbf{q}_0)\rangle=(|1\rangle-\cos{\theta_2}|2\rangle+\sin{\theta_2}|4\rangle)/\sqrt{2}$.  
    Figure~2\textbf{b} shows the dependance of the final state on $\theta_2$ for both ramp directions.
    The control vector $\mathbf{q}$ traces out a closed loop $\mathcal{C}$ in parameter space with $\mathbf{q}(t)=(-\varOmega_{\mathrm B},-\varOmega_{\mathrm A}\cos(2\mathrm{\uppi} t/T),-\varOmega_{\mathrm A}\sin(2\uppi t/T),0,0)$, a circle in 5-space subtending a solid angle $\varXi=2\uppi(1-\sin\theta_2)$ with respect to the origin.
    In the following sections, we will see that the states resulting from the $\pm$ ramps can be related by process tomography and WL measurement.

	\textit{Quantum process tomography} 
    We fully characterize the process of W.-Z. phase acquisition using quantum process tomography~\cite{Nielsen&Chuang, MLE} within the ground DS.
    An arbitrary transformation (operation) on a quantum system with initial density operator 
    $\hat \rho_{\rm ini}$ can be described by the action of Kraus operators $\hat{K}_k$:
    $\hat{\rho}_{\rm fin} = \sum_{k} \hat{K}_k\hat{\rho}_{\rm ini} \hat{K}_k^{\dagger}$. 
    The Kraus operators $\hat{K}_k$ completely describe the whole process, and can be expanded by a basis for operators $\{\hat{E}_i\}$ as $\hat{K}_k=\sum_i c_{ki} \hat{E}_i$, where $c_{ki}(\in \mathbb{C}$) is the coefficient.
    Thus, the density operator encoding the state within the DS transforms as
    $\hat{\rho}_{\rm fin}=\sum_{i,j} \hat{E}_i \hat{\rho}_{\rm ini} \hat{E}_j^{\dagger} \chi_{ij}$,
    with weights given by the process matrix $\chi_{ij}=\sum_k c_{ki} c_{kj}^{*}$.
    The process matrix $\chi$ completely and uniquely represents arbitrary transformations.
     {In our experiment, the path-dependent process matrix $\chi$ describes the transformation from the initial quantum state at $\mathbf{q}(t=0)=\mathbf{q}_0$ to the final state at $\mathbf{q}(t=T)=\mathbf{q}_0$, characterizing the W.-Z. phase acquisition process including any potential experimental imperfection.}
     {Under ideal unitary evolution}, each element $\chi_{ij} = \tr(\hat{U} \hat{E}_i) \tr(\hat{U} \hat{E}_j)^{*}/4$ is derived from the non-Abelian W.-Z. phase . 

    We experimentally obtain the process matrix $\chi$ by repeating the measurement illustrated in Fig.~2 for four-independent initial states
    $(|\mathrm{A}\rangle=|{\uparrow}_{-}(\mathbf{q}_0)\rangle$, 
    $|\mathrm{B}\rangle=|{\downarrow}_{-}(\mathbf{q}_0)\rangle$, 
    $|\mathrm{C}\rangle=(|\mathrm{A}\rangle+|\mathrm{B}\rangle)\sqrt{2}$, 
    $|\mathrm{D}\rangle=(|\mathrm{A}\rangle+\mathrm{i}|\mathrm{B}\rangle)\sqrt{2})$
    and applying maximum likelihood estimation to obtain a positive semi-definite and Hermitian matrix $\chi$.
     {We took $\{\hat{E}_i\}=\{\hat{I}_0, \hat{\sigma}_{\rm x}, \hat{\sigma}_{\rm y}, \hat{\sigma}_{\rm z}\}$ as the basis.}

    Figure~3\textbf{a}, \textbf{b} illustrates the reconstructed process matrices of the non-Abelian W.-Z. phase obtained for the forward and the reverse ramps at $\theta_2=\uppi/4$.
    The two results for opposite ramps along the same path show that the real part of $\chi$ takes almost the same values, whereas the imaginary parts of $\chi$ have the opposite sign.
    This trend can be explained from the definition of the W.-Z. phase (Eq.~\eqref{holonomy}) satisfying the relation $\hat{U}_{\mathcal{C}_+} = \hat{U}_{\mathcal{C}_-}^\dagger$ and thus $\chi_{ij}({\mathcal{C}_+})=\chi_{ij}^{*}({\mathcal{C}_-})$ for the process matrices of the non-Abelian W.-Z. phase.
    The above behavior of the process matrix elements holds for different coupling ratios (i.e. $\theta_2$) as shown in Fig.~3\textbf{c}, where different non-Abelian W.-Z. phases are realized (See Methods).
    The result, which is in stark contrast to the Abelian Berry phase, is in excellent agreement with the generalized relation for holonomy.
    
    The process matrix allows us to evaluate the fidelity of our holonomic control within the DS.
    Using the analytical expression for the non-Abelian holonomy $\hat{U}_\mathcal{C}$, the fidelity of the process shown in Fig.~3 \textbf{a}, \textbf{b} reached as high as $\mathcal{F_C}_{+}=0.98$ for forward ramp and $\mathcal{F_C}_{-}=0.96$ for reverse ramp even for finite ramp time. Here the fidelity is defined as $\mathcal{F}=\tr(\chi_{\rm th}\chi)$, where $\chi_{\rm th}$ is theoretical expected process matrix.
    This high fidelity then enabled us to characterize the W.-Z. phase with high accuracy.
    It has been argued that the non-adiabatic effect does not contribute to the state evolution in the DS up to first order, even though the state deflects from the adiabatic limit~\cite{Qi2016, YangMonopole2018}.
    
\textit{Wilson loop} 
    The above measurements depend on a choice of basis, i.e., of gauge, whereas the WL does not.
	The absolute value of the WL is $|W_\mathcal{C}|=2\sqrt{\chi_{00}}$, derived from a single component of the process matrix shown in Fig.~3.
    Figure~4 shows $|W_\mathcal{C}|$ for forward ($\mathcal{C}_+$) and reverse ramps ($\mathcal{C}_-$) as the path $\mathcal{C}$ is varied by changing the solid angle $\varXi(\theta_2)$.
    The expected relation $|W_{\mathcal{C}_+}|=|W_{\mathcal{C}_-}|$ is evidenced in experimental data, which also shows good agreement with the analytical curve for adiabatic control. 
    At $\varXi=2\uppi$ ($\theta_2=0$) our system decomposes into two uncoupled two-level systems; the geometric phase is $\hat{U}_{\mathcal C}=-\hat{I}_0$ and $W_{\mathcal{C}_\pm}=-2$ results from the $\pm\uppi$ Berry phase of these two-level systems seperatly.
    The experimental result shows the maximum change of the possible $|W_\mathcal{C}|$ (from 0 to 2), manifesting the non-Abelian nature of the geometric property.
    Since the system is time-reversal (TR) invariant~\cite{footnote2},
    the global phase factor should be $\pm n \uppi$ ($n$ is an integer) making the WL real valued.
    Stokes' theorem, which  equates Berry phase to the enclosed Berry curvature, is valid for the Abelian case, but not for the non-Abelian case.
    Still, for closed circular paths on a hypersphere centered at the monopole, the WL is $W_{\mathcal{C}_\pm}=2\cos(\varXi/2)$, determined by the solid angle subtended.
    Like the Aharonov-Bohm phase from a Dirac monopole, equal to half the solid angle, the angle dependence of the WL characterizes the Yang monopole (but is not proportional to the solid angle, as would be implied by Stokes' theorem).
     
    We gained further insight to the WL using the eigenvalues of the W.-Z. phase factor obtained from our measurements. 
    Since holonomy $U_\mathcal{C}$ is unitary, its unit-magnitude eigenvalues $\exp(\mathrm{i}\lambda_j)$ are given by the arguments $\lambda_1$ and $\lambda_2$.
     This immediately relates the WL to the gauge-independent difference $\updelta\uplambda=|\lambda_1-\lambda_2|$ via $|W_\mathcal{C}|=2 |\cos(\updelta\uplambda/2)|$.
	Figure ~4 \textbf{b} shows the phase difference inferred from WL measurements in good agreement with the theory.
	The inset illustrates that for our TR invariant system, the real-valued WL implies $\lambda_1=-\lambda_2$ and shows that the WL directly provides the eigenvalues of W.-Z. phase factor up to a $n \uppi$ phase uncertainty.

\section{Discussion}
    Our experiment realized Wu and Yang's gedanken experiment~\cite{Wu1975} to apply the generalized Aharonov-Bohm effect to the ${\rm SU}(2)$ isospin doublet of neutron and a proton, as an isospin gauge field detector.  Such experiments remain impractical for probing the standard model's combined ${\rm U}(1)\times {\rm SU}(2) \times {\rm SU}(3)$ gauge symmetry, but further progress in quantum analogues such as ours can shed light on the operation of such experiments.  An exciting next step in this direction would be creating a monopole source of a ${\rm SU}(3)$ gauge field (requiring a three-fold degenerate manifold), in analog to the Dirac monopole's ${\rm U}(1)$ gauge field (for a single non-degenerate state) and the Yang monopole's ${\rm SU}(2)$ gauge field (the two-fold degenerate manifold discussed here).

    Our experiments demonstrated essentially the full set of high-fidelity SU(2) holonomic control in a subspace which was protected against environmental noise and imperfections.  
	In the Bloch sphere picture, the process can be regarded as holonomic single qubit gate operation~\cite{Duan2001}, 
    where the Bloch vector is rotated by an angle of $\pm2\uppi \sin\theta_2$ around an axis $(-\cos\theta_2, 0, \sin\theta_2)$ depending on the path $\mathcal{C}_\pm$. 
    Universal operation are possible with more general path for Eq.~\ref{Hamiltonian},
     {since there is no experimental limitation in our implementation}.
    We note that our four level system can be used to code two qubits simultaneously -- one per degenerate manifold -- this may have application for redundant encoding, or possibly even independent holonomic control.
    
    This scheme forms a building-block broadly applicable to a wide range of systems including trapped ions, superconducting qubits~\cite{Hauke2012}, NV centers and other solid-state spins.
    Applications in this broader setting include precision measurement (e.g. magnetometry~\cite{Arai2018}), quantum gate operations, and quantum simulation using adiabatic W.-Z phases.

\section{Methods}
\subsection{\normalsize{Atom preparation and atom number counting}}

    Bose-Einstein Condensates (BECs) of rubidium-87 of $\approx 1 \times 10^5$ were prepared in a crossed optical dipole trap formed by two horizontal 1064 nm optical trapping beams
    with the trapping frequencies $(f_{\rm x}, f_{\rm y}, f_{\rm z}) \approx (50, 110, 70)$ Hz, 
    where the y-axis is along the direction of gravity.
    Initially, the BECs were prepared in the $|1,-1\rangle$ state.
    Atoms were then transferred to prepare a superposition state of $|1,0\rangle$ and $|2,0\rangle$
    by rf and microwave pulses, 
    which is the ground state of our Hamiltonian in Eq.~\eqref{Hamiltonian}
    at $\mathbf{q}_\mathrm{N} {=|\mathbf{q}_\mathrm{N}|(0,0,0,1,0)}$.
    The bias magnetic field of 19.8 G pointing along the z-axis was stabilized for long term drift
    at 2.5 ppm.
    
    We performed an absorption imaging and Stern-Gerlach measurements
    to resolve the atoms in the hyperfine ground states.
    After the rf and microwave control was finished,
    we abruptly turned off the optical dipole trap beams for time-of-flight (TOF).
    During the TOF, a magnetic field gradient pulse was applied to perform Stern-Gerlach measurement,
    which separated 
    atoms in $|1,0\rangle$ and $|2,0\rangle$ from those in $|1,1\rangle$ and $|2,-1\rangle$
    in space.   
    We imaged the atoms in $F=2$ manifold by illuminating a probe pulse resonant to 
    $5S_{1/2}, F=2 \rightarrow 5P_{3/2}, F=3$ 
    transition after TOF of 23.2 ms.
    A short repump laser pulse resonant wit the $5S_{1/2}, F=1 \rightarrow 5P_{3/2}, F=2$ transition 
    was applied before the probe pulse in order to image atoms in $F=1$ and $F=2$ manifolds.
    When we focused on the state in ground DS, 
    we apply a $\uppi$-pulse resonant with the microwave transition 
    $|1,0\rangle \leftrightarrow |2,1\rangle$
    to swap the population between the two states right before the TOF
    and only measure atoms in the $F=2$ manifold.
    This allowed us to measure the relative population 
    $(N_{\uparrow}-N_{\downarrow})/(N_{\uparrow}+N_{\downarrow})$
    with a single shot image.
    Here $N_{\uparrow} (N_{\downarrow})$ is the atom number in 
    $|{\uparrow}\rangle=|1,0\rangle (|{\downarrow}\rangle=|2,0\rangle)$ state
    before the microwave $\uppi$-pulse was applied.

\subsection{\normalsize{The Dirac matrices}}
    As the representation of the Dirac matrices, we take
    $\hat{\varGamma}_1=\hat{\sigma}_{\rm y} \otimes \hat{\sigma}_{\rm y}$,
    $\hat{\varGamma}_2=\hat{I}_0 \otimes \hat{\sigma}_{\rm x}$,
    $\hat{\varGamma}_3=-\hat{\sigma}_{\rm z} \otimes \hat{\sigma}_{\rm y}$,
    $\hat{\varGamma}_4=\hat{I}_0 \otimes \hat{\sigma}_{\rm z}$
    and $\hat{\varGamma}_5=\hat{\sigma}_{\rm x} \otimes \hat{\sigma}_{\rm y}$.
    Here $\hat{\sigma}_i, (i=x,y,z)$ are the Pauli operators,
    $\hat{I}_0$ is the identity operator,
    and $\otimes$ is the Kronecker product.
     {The left (right) Pauli operators in the products operate on the $F=\{0,1\}$ ($|m_F|=\{0,1\}$) space.}
    Each Dirac matrix has eigenvalues of $\pm 1$, each of which is two-fold degenerate.

\subsection{\normalsize{Pulse control for state preparation and state mapping}}
     For state preparation and mapping, we applied rf and microwave pulses with the same coupling configuration as in Fig.~1 (b). The unitary operator corresponding to these operations can be expressed using the following time-independent Hamiltonian.
        \begin{equation}
            \hat{H}_{\rm map}(\mathbf{q})=-\frac{\mathrm{i}\hbar}{2} (I_0 \otimes \sigma_{\rm z})(q_1 \hat{\varGamma}_1+q_2 \hat{\varGamma}_2+q_3 \hat{\varGamma}_3+q_5 \hat{\varGamma}_5).
        \end{equation}
    Note that only the relative phases of the cyclic coupling are different from the Hamiltonian in Eq.~\ref{Hamiltonian} with zero detuning ($q_4$=0).
    The unitary evolution during the pulsing is then
       \begin{equation}
            \hat{U}_{\rm trans}(t,\mathbf{q})=\exp(-\mathrm{i} \hat{H}_{\rm map}(\mathbf{q}) t),
        \end{equation}
    where $t$ is the pulse duration, and the state oscillates at a period determined by the energy gap ($\Updelta\mathrm{E}$).
    For $t_{\rm prep}=\uppi/(2\sqrt{\varOmega_{\rm A}^2+\varOmega_{\rm B}^2})$, the basis states at $\mathbf{q}_{\rm N}$, 
    which are $|{\uparrow}\rangle=|1\rangle$ and $|{\downarrow}\rangle=|3\rangle$, are mapped to $|{\uparrow}_{ {-}}(\mathbf{q}_0)\rangle$ and $|{\downarrow}_{ {-}}(\mathbf{q}_0)\rangle$ at $\phi_{\rm A}=\phi_{\rm B}=0$, respectively.
    For $t_{\rm map}=3\uppi/(2\sqrt{\varOmega_{\rm A}^2+\varOmega_{\rm B}^2})$, the basis states at $\mathbf{q}$ along $\mathcal{C}$,
    which are $|{\uparrow}_{ {-}}(\mathbf{q})\rangle$ and  {$|{\downarrow}_{ {-}}(\mathbf{q)})\rangle$},
    are mapped back to $|{\uparrow}\rangle$ and $|{\downarrow}\rangle$, respectively.
    The former pulse operation ($\hat{U}_{\rm trans}(t_{\rm prep},\mathbf{q}_0)=\hat{U}_{\rm prep}$) is applied for preparing 
    the eigenstates at $\mathbf{q}_0$ before the phase ramp,
    whereas the latter ($\hat{U}_{\rm trans}(t_{\rm map},\mathbf{q})=\hat{U}_{\rm map}$) is applied 
    after the phase ramp along the loop $\mathcal{C}$ to read out the state.

\subsection{\normalsize{The basis states for the DS}}
    We take the following eigenstates for the basis of the ground DS at $\mathbf{q}$
    for the region in parameter space we have experimentally explored
    $(\delta=0, \phi_{\rm B}=0, \phi_{\rm A}\in[0,2\uppi],\theta_2\in[0,\uppi/2])$,    
        \begin{equation}
        \begin{split}
            |{\uparrow}_-(\mathbf{q})\rangle&=(|1\rangle-\mathrm{e}^{-\mathrm{i}\phi_{\rm A}}\cos{\theta_2}|2\rangle+\sin{\theta_2}|4\rangle)/\sqrt{2}\\
            |{\downarrow}_-(\mathbf{q})\rangle&=(-\sin{\theta_2}|2\rangle+|3\rangle-\mathrm{e}^{\mathrm{i}\phi_{\rm A}}\cos{\theta_2}|4\rangle)/\sqrt{2}
        \end{split}
        \label{basisstate}
        \end{equation}
    Using the basis states, the pure state within the DS at $\mathbf{q}$ is described by
    \begin{equation}
        |\varPsi(\mathbf{q})\rangle=c_{\uparrow} |{\uparrow}_-(\mathbf{q})\rangle + c_{\downarrow}|{\downarrow}_-(\mathbf{q})\rangle,    
    \end{equation}
    which can be represented by a two-component spinor
    $\varPsi=(c_{\uparrow}, c_{\downarrow})^{\rm T}$,
    where $|c_{\uparrow}|^2+|c_{\downarrow}|^2=1$ is met.
    
    Each eigenstate can be prepared by applying the cyclic coupling pulse as described above to one of the bare spin states.
    \begin{equation}
    \begin{split}
        |{\uparrow}_-(\mathbf{q})\rangle &= \hat{U}_{\rm trans}(t_{\rm prep},\mathbf{q})|{\uparrow}\rangle,\\
        |{\downarrow}_-(\mathbf{q})\rangle &= \hat{U}_{\rm trans}(t_{\rm prep},\mathbf{q})|{\downarrow}\rangle,
    \end{split}
    \end{equation}
    where $|{\uparrow}\rangle=|1\rangle$ and $|{\downarrow}\rangle=|3\rangle$
    is the basis state of the ground DS at $\mathbf{q}_{\rm N}$. 
    Therefore, the four initial eigenstates 
    ($|\mathrm{A}\rangle$, $|\mathrm{B}\rangle$, $|\mathrm{C}\rangle$ and $|\mathrm{D}\rangle$)
    at $\mathbf{q}_0$ can be prepared by applying the pulse for the duration $t_{\rm prep}$ with the parameter vector $\mathbf{q}_0$ to the states  
    $|{\uparrow}\rangle$,  $|{\downarrow}\rangle$,
    $(|{\uparrow}\rangle+|{\downarrow}\rangle)/\sqrt{2}$,
    and $(|{\uparrow}\rangle+\mathrm{i}|{\downarrow})/\sqrt{2}$, respectively.
    For the state mapping, the basis states of the DS
    at $\mathbf{q}$ can be mapped to the bare spin states. 
        \begin{equation}
        \begin{split}
            |{\uparrow}\rangle &= -\hat{U}_{\rm trans}(t_{\rm map},\mathbf{q})|{\uparrow}_-(\mathbf{q})\rangle,\\
            |{\downarrow}\rangle &= -\hat{U}_{\rm trans}(t_{\rm map},\mathbf{q})|{\downarrow}_-(\mathbf{q})\rangle,
        \end{split}
        \end{equation}

\subsection{\normalsize{Quantum state tomography}}
    After the state acquired the W.-Z phase, we measured the final state within the DS
    by evaluating the Bloch vector
    $(\langle{\hat{\sigma}_{\rm x}(\mathbf{q})}\rangle,\langle{\hat{\sigma}_{\rm y}(\mathbf{q})}\rangle,
    \langle{\hat{\sigma}_{\rm z}(\mathbf{q})}\rangle)$.
    Here the Pauli operators are defined from the basis states of the DS at $\mathbf{q}$.
    
    Using the state mapping procedure described above, 
    the target Bloch vector is obtained by performing state tomography for
    the superposition states in the microwave clock transition
    ($|1,0\rangle \leftrightarrow |2,0\rangle$).
    The z-component was obtained from the population imbalance
    $(N_{\uparrow}-N_{\downarrow})/(N_{\uparrow}+N_{\downarrow})$.
    The x or y-component was obtained by rotating the Bloch vector with a $\uppi/2$-pulse with an appropriate microwave phase before measuring the population imbalance.

\subsection{\normalsize{Quantum process tomography using maximum likelihood estimation}}
     {
    Quantum process tomography is a scheme to characterize unknown quantum process
    by the knowledge of final output states for different input states.
    This information is used to reconstruct the process matrix that characterizes an arbitrary transformation.
    To determine $\chi$, $d^2$ linearly independent input states are required for a $d$-dimensional Hilbert space. For our DS with $d=2$, four inputs states, $|\mathrm{A}\rangle, |\mathrm{B}\rangle, |\mathrm{C}\rangle, |\mathrm{D}\rangle$ are taken as a set of inputs.
    }
    In order to the find physical process matrix $\chi$ that
    represents the W.-Z. phase from our measurement,
    we adopted maximum likelihood estimation in the quantum process tomography.
    For the process matrix to be physical, we define $\chi$ as
    \begin{equation}
        \chi=T^\dagger T/\tr(T^\dagger{T}),
    \end{equation}
    where $T$ is the lower triangular matrix of the form
    \begin{equation}
        T=
        \begin{bmatrix}
            t_1 & 0 & 0 & 0\\
            t_5 +\mathrm{i} t_6 & t_2 & 0 & 0\\
            t_{11} +\mathrm{i} t_{12} & t_7+\mathrm{i} t_8 & t_3 & 0\\
            t_{15} +\mathrm{i} t_{16} & t_{13} +\mathrm{i} t_{14} & t_9 +\mathrm{i} t_{10} & t_4
        \end{bmatrix}
    \end{equation}        
    where $t_{i}, (i=1,...,16)$ is real.
    We define a minimizing function $f(\mathbf{t})$ as
        \begin{equation}
            f(\mathbf{t})=\sum_{j} \left[
            \hat{\rho}_{{\rm fin},j}-\left\{
            \sum_{m,n}\hat{E}_m \hat{\rho}_{\rm ini} \hat{E}_n^\dagger \left(\frac{T^\dagger T}{\tr(T^\dagger{T})}\right)_{mn}
            \right\}_j \right]^2,
       \end{equation}
    where $\mathbf{t}=(t_1,t_2,\cdots, t_{16})$,
    $j\in\{{\rm A,B,C,D}\}$ distinguish the four initial states in the W.-Z. phase measurements,
    $\hat{\rho}_{\rm ini}=|j \rangle \langle j|$, 
    and $\hat{\rho}_{\rm fin}$ is the density operator for the state after it traced out the (open or closed) 
    loop $\mathcal{C}$.
    An average of measurements was used for each $\hat{\rho}_{{\rm fin},j}$.
    We numerically minimize $f(\mathbf{t})$ for the parameter vector $\mathbf{t}$ 
    to obtain optimum $T$ and the process matrix $\chi$.

\subsection{\normalsize{Synthetic non-Abelian SU(2) gauge field and Wilson loop}}
    Consider a quantum system with a Hamiltonian $\hat{H}(\mathbf{q})$ 
    that depends continuously on the position vector $\mathbf{q}=(q_1,q_2,\dots)$.
    The system is described by eigenstates and eigenenergies 
    \begin{equation}
        \hat{H}(\mathbf{q})|\varPsi_{n\alpha}(\mathbf{q})\rangle = E_n (\mathbf{q})|\varPsi_{n\alpha} (\mathbf{q})\rangle, 
    \end{equation}
    where $|\varPsi_{n\alpha} (\mathbf{q})\rangle$ $(\alpha=1,2,\dots,N_\alpha)$ is 
    $N_\alpha$-fold degenerate eigenstate with energy $E_n$ forming an $N_\alpha$-fold DS.
    For quantum states in a single energy level $E_n$, a gauge potential called the Berry connection
       \begin{equation}
            \mathcal{A}_{q_m}^{\alpha\beta}(\mathbf{q})=\mathrm{i}\langle\varPsi_\alpha(\mathbf{q})|\partial/\partial q_m|\varPsi_\beta(\mathbf{q}) \rangle,
        \end{equation}     
    is encoded in the systems’ eigenstates, 
    where $ \mathcal{A}_{q_m}^{\alpha\beta}$ is the $m$-th component of the vector gauge field $\boldsymbol{\mathcal{A}}$ 
    represented as $N_\alpha$-by-$N_\alpha$ matrix. Here, we omitted $n$ in the l.h.s. for simplicity, and the matrix indices take $\alpha,\beta \in \{1,2,\cdots, N_\alpha\}$.
    The gauge field (Berry connection) is non-Abelian when two components of the gauge field do not commute with each other.

    Now, we consider the gauge field for the Hamiltonian in Eq.~\eqref{Hamiltonian}. 
    We focus on the parameters relevant to the experiment ($\varOmega_{\rm A}=\varOmega\cos\theta_2,\varOmega_{\rm B}=\sin\theta_2,\delta=0$, and $\phi_{\rm B}=0$).
    The non-Abelian SU(2) Berry connection for the two-fold degenerate ground states is
       \begin{widetext}
       \begin{equation}
        \begin{split}
            \mathcal{A}_{\upphi_{\rm A}}(\mathbf{q})=\frac{1}{2}
            \begin{pmatrix}
                \cos{^2}\theta_2 &
                \mathrm{e}^{\mathrm{i}\phi_{\rm A}}\sin{\theta_2}\cos{\theta_2} \\
                \mathrm{e}^{-\mathrm{i}\phi_{\rm A}}\sin{\theta_2}\cos{\theta_2} &
                -\cos^{2}\theta_2 \\
            \end{pmatrix}
            \\
            =(\cos{\phi_{\rm A}}\sin{\theta_2}\cos{\theta_2}\sigma_{\rm x}-\sin{\phi_{\rm A}}\sin{\theta_2}\cos{\theta_2}\sigma_{\rm y} +\cos{^2}\theta_2 \sigma_{\rm z})/2.
        \end{split}
        \end{equation}
        \end{widetext}
    
    From the definition in Eq.~\eqref{holonomy},
    we obtain W.-Z. phases factors and Wilson loops (WLs) for the paths $\mathcal{C}_\pm$,
        \begin{widetext}
        \begin{equation}
            U_\mathcal{C_\pm}=
            \begin{pmatrix}
                -\cos(\uppi\sin\theta_2) \pm \mathrm{i}\sin\theta_2\sin(\uppi\sin\theta_2)&
                \mp \mathrm{i}\cos\theta_2\sin(\uppi\sin\theta_2)\\
                \mp \mathrm{i}\cos\theta_2\sin(\uppi\sin\theta_2)&
                 -\cos(\uppi\sin\theta_2) \mp \mathrm{i}\sin\theta_2\sin(\uppi\sin\theta_2)\\
            \end{pmatrix}
            .
          \end{equation}
        \end{widetext}    
        \begin{equation}
             W_\mathcal{C_\pm}=-2\cos(\uppi\sin\theta_2).        
        \end{equation}
    The physical process can be regarded as holonomic single qubit gate operation, 
    where the two eigenstates of the degenerate level are taken as the qubit basis states and the Bloch vector representing the qubit is rotated by an angle of $\pm2\uppi \sin\theta_2$ around an axis $(-\cos\theta_2, 0, \sin\theta_2)$.
    The dependence on the ramp direction for the W.-Z. phase factors
    ($\hat{U}_{\mathcal{C}_+}=\hat{U}^\dagger_{\mathcal{C}_-}$),
    and the WLs
    ($W_{\mathcal{C}_+}=W^*_{\mathcal{C}_-}$) can be clearly seen.
    Both the W.-Z. phases factor and the WL do not depend on $\varOmega$,
    thus they are robust against fluctuation in the coupling strength. 
    By varying the rf phase $\phi_{\rm A}$, the SU(2) WL covers the full range
    ($-2 \leq W_{\mathcal{C}}\leq 2$), realizing various non-Abelian SU(2) holonomic controls.    
    
\subsection{\normalsize{Wilson line for an open path (theory)}}
    In the following we give an argument on non-cyclic W.-Z. phase and Wilson line for an open path.
    The definition for  non-cyclic W.-Z. phase and Wilson line are essentially the same as the cyclic case, except the integral is taken over for an open path $\mathcal{C}$. 
        \begin{equation}
            W_\mathcal{C}= \tr(\hat{U}_\mathcal{C})
            = \tr\left[\mathcal{P}\exp\left(\mathrm{i} \int_\mathcal{C} \boldsymbol{\mathcal{\hat{A}}}_\mathbf{q} \cdot d\mathbf{q}\right)\right],
        \end{equation}
    where $\boldsymbol{\mathcal{\hat{A}}}$ is non-Abelian Berry connection.
    
    Consider a spinor state vector $|\varPsi\rangle$ representing the state within the degenerate subspace (DS). Under local gauge transformation, the wavefunction transform as
    \begin{equation}
       |\varPsi\rangle \rightarrow \hat{V}(\mathbf{q}) |\varPsi\rangle, 
    \end{equation}   
    where $\hat{V}(\mathbf{q})$ is a position-dependent unitary operator. 
    This can be regard as a change in the basis states for the DS.
    Accordingly, the non-cyclic W.-Z. phase factor transforms as
        \begin{equation}
              \hat{U} _\mathcal{C}   \rightarrow
             \hat{V} (\mathbf{q}_{\rm f}) \hat{U} _\mathcal{C} \hat{V} ^\dagger(\mathbf{q}_0),   
    \end{equation}
    where $\mathbf{q}_0$ and $\mathbf{q}_{\rm f}$ are the start point and end point of the open path $\mathcal{C}$, respectively. 
    Manifestly, the r.h.s depends on the unitary operators, 
    $\hat{V} (\mathbf{q}_{0})$ and $\hat{V} (\mathbf{q}_{\rm f})$.
    When the trace is closed ($\mathbf{q}_{0}=\mathbf{q}_{\rm f}$), the Wilson line 
    is equivalent to the WL and is gauge-independent.

    For our experimental parameters for Wilson line measurement in Fig.~5 
    ($\delta=0$, $\phi_{\rm B}=0$ and $\theta_2=\uppi/4$),
    the non-cyclic non-Abelian W.-Z phase and Wilson lines along a segment 
    {} are
    \begin{widetext}
    \begin{equation}
            U_\mathcal{C}=
        \begin{pmatrix}
            \mathrm{e}^{\mathrm{i}\phi/2}\left[\cos \left(\frac{\phi}{2\sqrt{2}}\right)-\frac{\mathrm{i}}{\sqrt{2}}
            \sin \left(\frac{\phi }{2 \sqrt{2}}\right)\right]&
            \mathrm{i} \mathrm{e}^{\mathrm{i}\phi/2}\sin \left(\frac{\phi }{2 \sqrt{2}}\right)/\sqrt{2}\\
            \mathrm{i} \mathrm{e}^{-\mathrm{i}\phi/2}\sin \left(\frac{\phi }{2 \sqrt{2}}\right)/\sqrt{2}&
            \mathrm{e}^{-\mathrm{i}\phi/2}\left[\cos \left(\frac{\phi}{2\sqrt{2}}\right)+\frac{\mathrm{i}}{\sqrt{2}}
            \sin \left(\frac{\phi }{2 \sqrt{2}}\right)\right]\\
        \end{pmatrix}
            ,
        \end{equation}
        \begin{equation}
         W_\mathcal{C} = 
         \sqrt{2} \sin \left(\frac{\phi }{2}\right) \sin \left(\frac{\phi }{2 \sqrt{2}}\right)+2 \cos \left(\frac{\phi }{2}\right) \cos \left(\frac{\phi }{2 \sqrt{2}}\right)
         .    
    \end{equation}
    \end{widetext}        
    Here we took the basis in Eq.~\eqref{basisstate} for the matrix representation.

    \begin{figure}
        \includegraphics[width=3.3in]{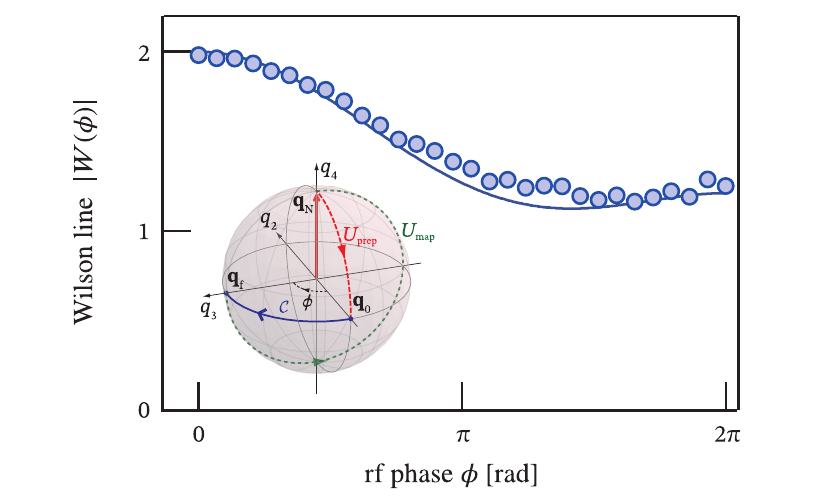}
        \caption{
        \textbf{{Wilson line.}}
        Absolute values of measured Wilson line $|W({\mathrm{\phi}})|$ 
        for open paths with variable path length characterized by rf phase range $\phi$. Theory curve (solid line) also shown.
        The inset illustrates the control sequence for Wilson line measurement 
        of a segment from $\mathbf{q}_0$ to $\mathbf{q}_{\rm f}$ on a circular loop.
        After preparing one of the eigenstates at $\mathbf{q}_0$, the rf phase $\phi_{\rm A}$ is ramped from $0$ to $\phi$.
        The red and green curves represent the pulse controls for
        the state preparation and the state mapping for the read-out. 
        The laboratory parameters are $\varOmega_{\rm A}=\varOmega_{\rm B}=h \times 1.4 $ kHz.
        }
    \end{figure}
    
\subsection{\normalsize{Wilson line for an open path (experiment)}}
    We show measurement on the Wilson line on open paths by observing non-cyclic W.-Z. phases.
    The non-cyclic W.-Z. phases is defined by simply replacing the closed path for
    the integral in Eq.~\eqref{holonomy}
    with an open path $\mathcal{C}$.
    The Wilson line, defined as the trace of the non-cyclic W.-Z. phase factor, is not gauge-independent, and thus it depends on the choice of the basis at both ends of the path. 	
    The experimental procedure is the same as the WL measurement, 
    except the rf phase ramp is halted at variable phase $\phi_{\rm A}=\phi$ ranging from $0$ to $2\uppi$.
    After preparing the eigenstates at $\mathbf{q}_0$, we ramp the control vector as
    $\mathbf{q}(t)=(-\varOmega_{\rm B}, -\varOmega_{\rm A} \cos(2\uppi t/T), -\varOmega_{\rm A} \sin(2\uppi t/T), 0, 0)$ for $t=[0, \phi T/2\uppi]$ until the control vector reaches $\mathbf{q}_{\rm f}=(-\varOmega_{\rm B}, -\varOmega_{\rm A} \cos{\phi}, -\varOmega_{\rm A} \sin{\phi}, 0, 0)$. 
    The final state within the DS at $\mathbf{q}_{\rm f}$ is always mapped to the DS at $\mathbf{q}=\mathbf{q}_0$
    for the state tomography.
    By performing the process tomography for the four-independent initial eigenstates, 
    the process matrix of the non-cyclic non-Abelian geometric phase is reconstructed in the same manner as in Fig.~3
	for each phase. 
    Figure~5 shows the obtained Wilson lines from the reconstructed process matrices
    for a choice of the basis states based on our experimental procedure.
    The Wilson line is trivially $W_\mathcal{C}=2$ at $\phi=0$ and becomes gauge-independent at $\phi=2\uppi$ where the trajectory is closed.
    
    The whole unitary process including the state preparation and the state mapping processes can be viewed as a local gauge transformation of the W.-Z phase: 
    $\hat{U}_\mathcal{C} \rightarrow  V(\mathbf{q}_{\rm f}) \hat{U}_\mathcal{C} V^{\dagger}(\mathbf{q}_0)=-\hat{U}_{\rm map}(\phi) \hat{U}_\mathcal{C}\hat{U}_{\rm prep}$,
    where $V(\mathbf{q})$ is a position-dependent unitary operator,
    and $\hat{U}_{\rm prep}$ ($\hat{U}_{\rm map}$) is a unitary operator that represents the pulse that maps the state within the DS at $\mathbf{q}_{\rm N}$ ($\mathbf{q}_{\rm f}$) to the state within the DS at $\mathbf{q}_0$. 
    This clearly illustrates that the Wilson line is gauge-independent only when $\mathbf{q}_{\rm f}=\mathbf{q}_0$, where it becomes equivalent to the WL.

    \begin{figure}
        \includegraphics[width=3.3in]{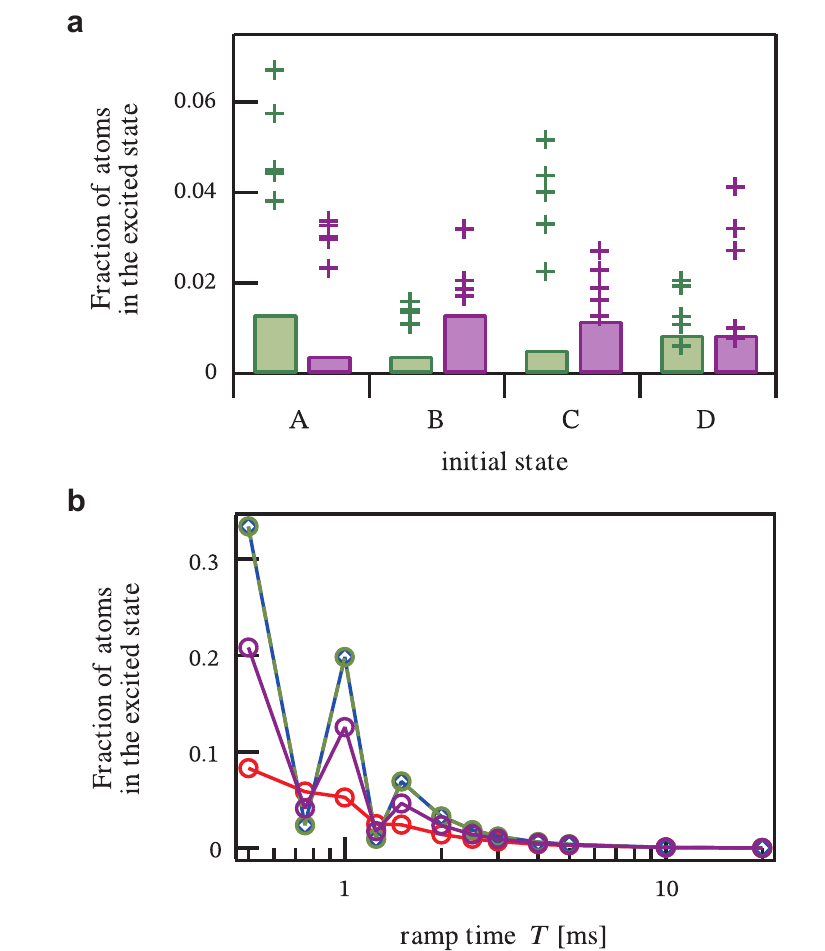}
        \caption{
        \textbf{{Excited state population after the W.-Z. phase acquisition.}}
        \textbf{a}
        Experimentally and numerically obtained excited state population at $T=2$ ms for $\theta_2=11\uppi/36$.
        Experimental data for the paths $\mathcal{C}_+$ (green points) and $\mathcal{C}_-$ (purple points)
        compared with numerical simulation for the paths $\mathcal{C}_+$ (green bar) and $\mathcal{C}_-$ (purple bar).
        \textbf{b} 
        Fraction of atoms in the excited state manifold after state acquired W.-Z. phase along $\mathcal{C}_-$ numerically simulated for the four initial states at $\theta_2=\uppi/4$.
        The four states are
        $|\mathrm{A}\rangle$ (red), $|\mathrm{B}\rangle$ (blue), $|\mathrm{C}\rangle$ (green), and $|\mathrm{D}\rangle$ (purple).
        Due to finite ramp time $T$ for tracing out the loop, the non-adiabatic effect is non-negligible when the ramp rate becomes comparable to the scale of the energy gap  
        ($\Updelta\mathrm{E} = h \times 2$ kHz).
        }
    \end{figure}
    
    \begin{figure}
        \includegraphics[width=3.3in]{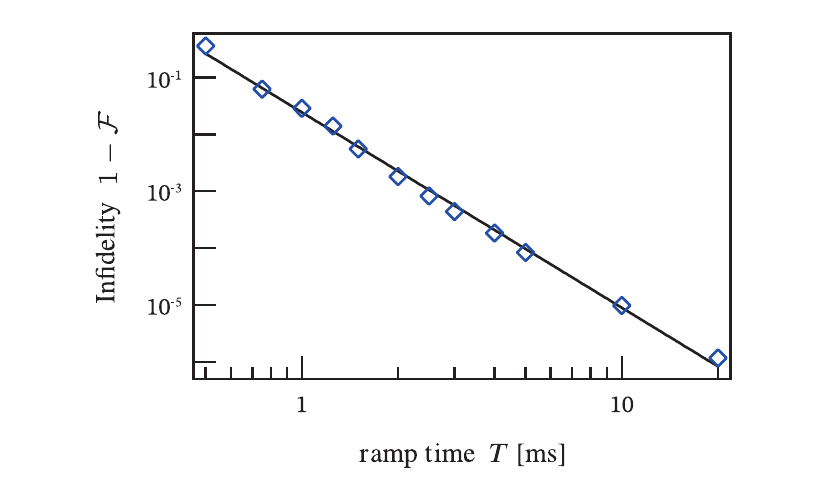}
        \caption{
        \textbf{{Infidelity of the non-Abelian W.-Z. phase ($1-\mathcal{F}$) 
        due to finite ramp time $T$.}}
        The non-Abelian W.-Z. phase factor is numerically evaluated by solving time-dependent Schr\"{o}dinger equation for our experimental condition for the path $\mathcal{C}_-$ with $\theta_2=\uppi/4$ and analyzed by following the same procedure used in the experiment to numerically obtain the process matrix. The solid line is a power-law fit to the numerical results.}
    \end{figure}

\subsection{\normalsize{Non-adiabatic effect due to finite ramp time}}\label{Non-adiabatic}
    
    Although we have focused on evolution within the ground-state manifold, 
    a small fraction of atoms can be populated to the excited state manifold due to the finite ramp time.
    We experimentally confirmed this by measuring the fraction of atoms in the excited state manifold. After the state mapping, we evaluated the fraction
    $N_{\rm e}/(N_{\rm e}+N_{\rm g})$, where $N_{\rm e}=N_{|2\rangle}+N_{|4\rangle}$ is the atom number in the excited state manifold, $N_{\rm g}=N_{|1\rangle}+N_{|3\rangle}$ is the atom number in the ground state manifold, and $N_{|i\rangle}$ is the atom number in the bare spin state $|i\rangle, (i=1,2,3,4)$. 
    The observed fraction of atoms, which depends on the initial state is negligibly small, and consistent
    with the numerical simulation (See Fig.~6\textbf{a}).
     {The dependence of the excited atomic fraction on the initial state can be understood by the state-dependent nature of the state deflection due to local non-Abelian gauge field~\cite{YangMonopole2018}.} 
    Longer ramp time led to a smaller fraction in the excited state manifold as confirmed by 
    the numerical simulation 
    [Fig.~6\textbf{b}]. 
    Experimentally, the fidelity of the our holonomic control is expected to be degraded for longer ramp time due to the small but finite energy gap opening in the nearly-degenerate levels, which we assume to be about 1\% of the energy gap of the system.

    Surprisingly, the fidelity in the W.-Z. phase measurement within the DS
    is robust against small non-adiabatic effect.
    Fig.~7 shows the numerically obtained fidelity of the W.-Z. phase 
    by varying the ramp time.
    For our experimental parameters with $\theta_2=\uppi/4$,
    the fidelity reaches 0.998\% at $T$=2 ms.

\subsection{\normalsize{Measurement of the energy gap}}\label{EnergyGap}

     The energy gap can be clearly measured by inducing coherent Rabi-like oscillations between the eigenstates.
     Figure~8 shows that the time evolution of the population imbalance $(N_{\rm e}-N_{\rm g})/(N_{\rm e}+N_{\rm g})$
     after abruptly turning on the cyclic coupling described by the Hamiltonian in Eq.~\eqref{Hamiltonian}.
     Since the system has only two eigenenergies, the state oscillates between the ground and excited eigenstates at the frequency determined by the energy gap.

    \begin{figure}
        \includegraphics[width=3.3in]{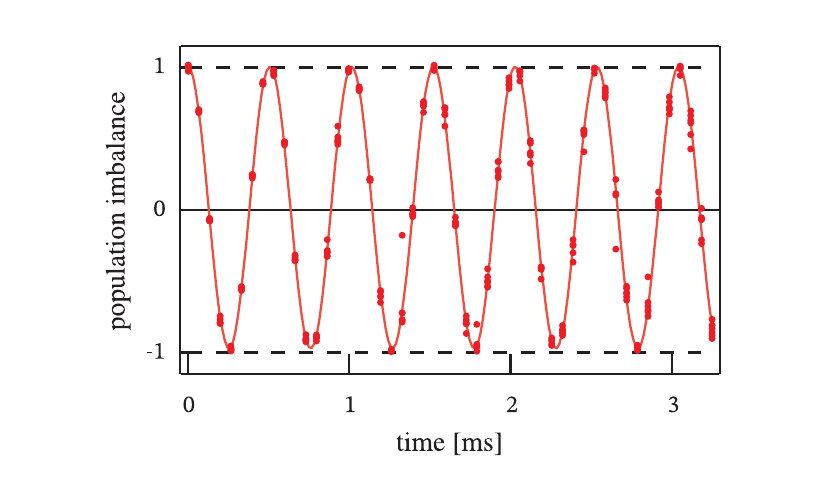}
        \caption{
        \textbf{{Rabi-like oscillation between the two eigenstates.}}
        The population imbalance $(N_{\rm g}-N_{\rm e})/(N_{\rm g}+N_{\rm e})$ was measured after the cyclic coupling for $\theta_2=\uppi/4$ was abruptly turned on with the BEC in  state $|2\rangle$. 
        The observed oscillation frequency of 2.0 kHz determines the energy gap of our system. 
        }
    \end{figure}
    
\section{Data availability}
The data are available from the corresponding author upon reasonable request.

\section{Acknowledgments}
\begin{acknowledgments}
    We acknowledge the support for this work provided by the AFOSRs Quantum Matter MURI, NIST and the NSF through the PFC at JQI. 
    S.S. acknowledges support from JST, PRESTO (JPMJPR1664) and JSPS (fellowship for research aboard). 
    We thank Mingwu Lu and Chris Billington for carefully reading our manuscript.
    
    This version of the article has been accepted for publication, after peer review but is not the Version of Record and does not reflect post-acceptance improvements, or any corrections. The Version of Record is available online at: http://dx.doi.org/10.1038/s41534-021-00483-2.
\end{acknowledgments}

\section{Competing Interests}
The authors declare no competing interests.

\section{Author Contributions}
    S.S., F.S.-C., Y.Y., and A.P. measured the data.
    S.S. conceived the experiment and analyzed the data. I.B.S. supervised the project. 
    All the authors contributed to discussion and preparation of the manuscript.

\section{Correspondence}
Seiji Sugawa (sugawa@ims.ac.jp) and I.B. Spielman (spielman@umd.edu)


\begin{thebibliography}{99}
\bibitem[35]{footnote1}
    The WL is not uniquely defined: in the condensed-matter literature the non-Abelian holonomy $\hat{U}$ is used synonymously with the WL, whereas in other contexts such as high-energy physics, $\tr(\hat{U})$ is associated with the WL. Both conventions are present in the quantum-gas literature, and we have adopted the latter definition.
\end{thebibliography}
\end{document}